\documentstyle[aps,prl,multicol,epsf]{revtex}

\def\ket#1{| #1\rangle}
\def\braket#1#2{\langle \, #1 \, | \, #2 \, \rangle}

\def\R{\hbox{\rm I \kern-5pt R}}

\title{Quantum Interrogation and the Safer X-ray}

\author{Adrian Kent${}^{1,2}$ and David Wallace${}^3$} 
\address{${}^1$ Hewlett-Packard Laboratories, Filton Road,\\
Stoke Gifford, Bristol BS34 8QZ, U.K.\\ 
${}^3$ Centre for Quantum Computation,  Clarendon Laboratory,\\
Department of Physics, University of Oxford, Parks Road,\\
Oxford OX1 3PU, U.K.
}
\date{February 2001} 

\begin{document} 
\maketitle
\begin{abstract}
We investigate quantum interrogation techniques which allow 
imaging information about semi-transparent objects
to be obtained with lower absorption rates than standard
classical methods.  We show that a gain proportional to 
$\log N$ can be obtained when searching for defects in 
an array of $N$ pixels, if it is known that at most $M$ 
of the pixels can have transparencies different from a  
predetermined theoretical value.  
A logarithmic gain can also be obtained when searching for 
infrequently occurring large structures in arrays.  
\end{abstract}
\vskip 5pt
${}^2$ On leave from DAMTP, University of Cambridge, 
Silver Street, Cambridge CB3 9EW, U.K. 

\begin{multicols}{2}

\section{Introduction} 

The well-known Elitzur-Vaidman quantum interrogation scheme\cite{ev} 
uses interferometry in a way which sometimes allows photo-sensitive 
objects to be detected without absorbing a photon.  
In EV's original example, starting from a mixed supply of 
``good'' bombs that will explode if a single photon hits the trigger and
of dud bombs which reflect photons without exploding, a 
guaranteed supply of unexploded good bombs can be produced. 
EV's probabilistic scheme was significantly refined by Kwiat 
et al.\cite{kwiatetal}, who showed that
the probability of absorption-free detection of
a photo-sensitive object can be made arbitrarily close to one
by a quantum Zeno interferometry technique. 

These schemes both distinguish perfectly opaque
from perfectly transparent (or, in the original formulation, perfectly 
reflecting) objects. 
They inspired the hope that effectively absorption-free quantum interrogation 
methods could also 
be used to discriminate  
semi-transparent objects of different transparencies.
Such absorption-free discrimination, if possible,  would be of great 
practical value.   It could, for example, allow a ``safe X-ray'', 
in which images of a patient's internal
structure could be obtained with arbitrarily low absorption of
harmful X-rays.  

In fact, though, the Kwiat et al. quantum Zeno technique does
not allow absorption-free discrimination between partially opaque
objects\cite{kwiatscripta}.   More generally, Mitchison and Massar
have shown\cite{mm} that there is a non-zero bound on the probabilities
of absorption involved in any scheme discriminating between 
two imperfectly opaque objects.  

The Mitchison-Massar bound might be 
taken as dampening hope of finding any seriously valuable 
imaging applications of quantum interrogation other than
discriminating a partially opaque and a completely
transparent object.  However, such pessimism would be unjustified.
We describe here two quantum interrogation schemes 
which show that quantum information
can indeed be used to extract imaging information about 
semi-transparent objects with significantly reduced 
absorption rates.  Though the safe X-ray is impossible, 
a somewhat safer X-ray is viable, at least in theory.  

We represent the object to be imaged as an array of 
$N$ pixels, labelled by $i$ from $1$ to $N$,
with transparencies $\alpha_i$ for the relevant 
radiation.  
That is, the interaction between a photon and the $i$-th pixel
is described by 
\begin{equation}
\ket{ \gamma^{\rm in}_i } \ket{i} \rightarrow \alpha_i  
\ket{ \gamma^{\rm out}_i } \ket{i} + 
\beta_i  \ket{ i^* }  \, , 
\end{equation} 
where $\ket{ \gamma^{\rm in}_i } $ and $\ket{ \gamma^{\rm out}_i}$ 
describe incoming and outgoing photons in a beam passing through
the $i$-th pixel, $\ket{i}$ is the initial state of the $i$-th pixel,
$\ket{ i^* }$ is the excited
state of the $i$-th pixel after absorbing a photon, and 
$ | \alpha_i |^2  + | \beta_i |^2 = 1$, with $ 0 \leq | \alpha_i | \leq 1 $. 

Our aim is to extract useful imaging information about the pixel array
transparencies
while minimizing the expected number of photons absorbed in the array. 
Specifically, we are interested in quantum methods which do better
in this respect than standard X-ray imaging.
A standard X-ray image is effectively obtained by firing a number 
of photons separately through each of a number of different pixels, 
counting the number of outgoing photons, and statistically inferring 
information about the individual pixel transparencies or some joint 
functions thereof.  This ``classical'' method makes no
essential use of the fact that quantum information is being 
generated in the imaging process.  Indeed, it makes no use of 
quantum theory at all, beyond the fact that photons are quantised.
Effectively the same method could be used with classical radiation, assuming
some finite bound on the sensitivity with which the amplitude of 
the outgoing radiation can be measured, and indeed this
is how most standard imaging schemes work in practice.   

From the perspective of quantum information theory, however, 
the classical method seems unlikely to be optimal for general problems.
The pixel array behaves like a non-unitary oracle in its action on 
incoming quantum states.  It is well known in quantum computing that
interrogating a unitary oracle by quantum superposition states 
and processing the outputs can be far more efficient than 
invoking it only on the basis states which define its action.   
It should not be surprising that this turns out also to be 
true of non-unitary oracles. 
In fact, as we show here, it is true in simple examples which arise
naturally as imaging problems.   

\section{Testing an array for defects} 

Consider the following situation.  We know that the transparency of 
pixel $i$ should be $\alpha_i$, if nothing untoward has
occurred.  However, we are concerned that there may be a small 
number of pixels whose transparency is in fact significantly different
from the theoretical value, and we wish to check on this.   To be precise, 
we have some fixed parameter $\epsilon$, and know that up to $M$ of 
the pixels may have an altered transparency $\alpha'_i$, 
where $| \alpha'_i - \alpha_i | \geq \epsilon $, while the 
rest of the transparencies are unaltered.   
We wish to be statistically certain, to within some prescribed degree
of confidence $(1- \delta)$, whether or not there are any 
altered transparency pixels.  If there are, we wish to identify them.   

This problem would arise, for example, for pixel arrays manufactured 
by a generally reliable process that occasionally suffers from 
some specific defect on individual pixels.  It is also an idealisation of the 
imaging problem that arises when we have previously taken an X-ray 
of a (then) healthy patient and now want to know 
whether any significant change has occurred, if we may assume
that any changes will be confined to a small part of the image. 

In the following discussion we take $M$, $\delta$ and $\epsilon$ to 
be fixed, with $N$ varying.  We are interested in the behaviour
for large $N$; in particular we assume $N$ is large compared to 
$M^2$, $\epsilon^{-2}$ and $\delta^{-1}$.  We also assume the 
$\alpha_i$ are generally not close to $0$ or $1$.  

The classical method requires estimating the
transparency of each pixel separately.  This is done by firing photons
through each pixel and ensuring that the transparencies can all be
identified, either as the expected $\alpha_i$ or as 
differing by at least $\epsilon$, with confidence $(1- \delta )$ 
that none of the $N$ identifications is erroneous.  
A classically obtained image giving this degree of confidence 
separately for each of the individual pixel 
identification requires $O( \epsilon^{-2} N )$ absorptions.    
However, as $N$ grows, the chance that at least one 
of the $N$ pixel transparencies will be
misestimated by more than $\epsilon$ grows logarithmically in $N$.  
Allowing for this, we see that the classical solution to the 
problem defined requires an absorption rate of $O( \epsilon^{-2} N \log (N) )$.

Consider now the following quantum method.  Each incoming photon
is split into an equally weighted superposition of states incident on 
each of the pixels.  That is:
\begin{equation} 
\ket{\gamma_{\rm in} } \rightarrow {1 \over {\sqrt{N}}} \sum_{i=1}^N 
\ket{\gamma_{\rm in}^i } \, .
\end{equation} 
Let the actual transparency of pixel $i$ be $\beta_i$. 
After passing through the pixels, the photon, if it was not absorbed, 
is in the state 
\begin{equation} 
 {C \over {\sqrt{N}}} \sum_{i=1}^N \beta_i \ket{ \gamma_{\rm out}^{i} } \, ,
\end{equation} 
where $C$ is a normalisation constant and $\ket{\gamma_{\rm out}^i }$ are
the states of photons emerging on the far side of pixel $i$.  

We now need to know whether the photon was absorbed or not.
If the absorptions themselves are not easily detected, this 
can in principle be tested by carrying out a von Neumann measurement
of the photon number of the outgoing beam state. 
Assuming no absorption, we now carry out a 
further von Neumann measurement, 
measuring the projection $P_{\psi}$ onto the state 
\begin{equation}
\ket{ \psi } = {C' } \sum_{i=1}^N \alpha_i \ket{
  \gamma_{\rm out}^i } \, . 
\end{equation} 
If we obtain eigenvalue $0$, we know that the normalised outgoing state
was not $\ket{\psi}$, and hence that not all the pixel transparencies 
can have the theoretical value.   
In this case, we apply a second measurement, simultaneously
measuring the projections onto the states $\ket{\gamma_{\rm out}^i}$ for 
$i$ from $1$ to $N$.  In other words, we look to see which of the
outgoing beams we find a photon in.  We take the answer $i$ to
indicate that pixel $i$ is defective.  

Reducing the probability of a false negative --- 
a defective pixel array producing
eigenvalue $1$ throughout the sequence of tests --- to 
$\delta$ requires $O( \log ( \delta^{-1}) N \epsilon^{-2} )$ such tests.
and hence (given our assumptions about the $\alpha_i$) this order
of absorptions.  
Comparing the classical technique, and considering only the
$N$-dependence for large $N$ (since the other parameters are
fixed), we find a reduction of a factor of $O( \log N )$ in the 
absorption rate.  

When the defect test is positive, the probability of mis-identifying a 
good pixel as defective when we measure the outgoing beam is 
$O ( M / N )$. 
If this level of certainty is acceptable, 
a pixel thus identified as defective can be excluded from later tests.  
Up to $M$ defective pixels can then be identified
as defective, with the probability of any erroneous identification 
being $O ( M^2 / N )$. 
If these certainty levels are insufficient, we may 
require more than one positive result to identify a defective pixel.  
For any fixed level of certainty, however, the overall gain over
the classical method remains proportional to $\log N$.   

These calculations
assume that the theoretical values of the $\alpha_i$ are determined
with absolute precision.   In the realistic case in which these
values are themselves subject to errors, of sizes bounded by
$\epsilon'$, the conclusions hold so long as 
$\epsilon' \ll (M \epsilon / N )$.    

\section{Searching for infrequently occurring large structures}  

We again consider an array of $N$ pixels, and suppose now that 
we are looking for a particular type of rarely occurring image.  
We model this as follows.  Our aim is to decide, with confidence
$(1 - \delta )$, whether the array has a
particular pattern $\{ \alpha_i^0 \}$ of transparencies.  
We suppose that the prior probability of this pattern occuring  
is some small but non-zero number $p$, and that with probability $(1-p)$ the
pixel transparencies are randomly drawn from identical
independent probability distributions.  To be definite, 
we suppose that in this latter case the pixel 
transparencies are uniformly distributed in the complex unit disc.   
We suppose also that $\exp ( - \sqrt{N} ) \ll \delta p  $.  

We neglect constant factors throughout this section, considering
only the degree of dependence on the parameters $N, \delta, p$.

Classically, we can only approach this problem by obtaining
statistical estimates of some or all of the $| \alpha_i |$, and
comparing the estimates to $| \alpha_i^0 | $.  
Estimating the $| \alpha_i | $ on $r$ pixels to within error $\epsilon$
requires $ \approx r \epsilon^{-2} $ photon measurements and absorptions.
Suppose that $r$ such measurements produce the estimates 
$ | | \alpha_i | -  | \alpha_i^0 | | < \epsilon$.  For this to give us
the required confidence that we have found an example of
the image requires that $\epsilon^{r} \approx \delta p$.
Minimizing with respect to $\epsilon$, we find this strategy
requires $\approx - \log ( \delta p )$ absorptions. 

In a quantum approach to the problem, we proceed as in the
previous example, preparing an equal superposition of 
photon states incident on the pixels and, if there is
no absorption, testing whether the emerging state is 
$\ket{ \psi^0 } = {C' } \sum_{i=1}^N \alpha_i^0 \ket{
  \gamma_{\rm out}^i }$.
In the case where the pixel transparencies were randomly drawn from
uniform distributions, the probability that they are such that the
emerging state $\ket { \psi }$ obeys $ | \braket{ \psi_0 }{ \psi } |^2 > 
{ 1 \over { \sqrt{N}} }$ is $\approx \exp ( - \sqrt{N} )$, which 
is negligible in our calculations.  So we may assume that a randomly
drawn array will have transparencies such that
 $ | \braket{ \psi_0}{ \psi } |^2 < { 1 \over { \sqrt{N}} }$. 
We can confidently conclude that the image sought is present
after $x$ successful (and no unsuccessful) tests, where 
$ ( {1 \over { \sqrt{N} } } )^x \approx \delta p $, so that 
$x \approx \max ( {{ - \log ( \delta p ) } \over { \log N }} , 1 )$.  
As this represents the order of the number of absorptions required,
we again find a logarithmic quantum advantage.  

These arguments generalise to a search
for an infrequently occurring image which may be any one of 
$N$ known possibilities whose transparencies define $N$ orthogonal
states $\ket{ \psi_i }$.  

\section{Conclusions} 

The methods we have described 
show that quantum interrogation can have useful advantages over
standard classical methods for realistic large array imaging problems. 
It is perhaps worth noting that these methods apply equally
well to time-dependent imaging problems.  A single array in our
model could, for example, represent a smaller array being repeatedly probed   
at a sequence of times.  

A further example of a problem in which there is a 
logarithmic quantum advantage has been found by Massar et al.\cite{mmp}. 
It would be good to have a general understanding of 
the range of problems in which there is a quantum advantage.     
It would also be useful to identify the advantage attainable
by optimal techniques, both for
the problems we have described and more generally.
Here the bounds obtained in independent work by  
Massar et al.\cite{mmp} may be of help.   

\section{Acknowledgements} 

We gratefully acknowledge discussions with Graeme Mitchison, and
thank him in particular for stimulating us to consider the
defective pixel identification step in our first quantum
interrogation example.   
AK was supported by a Royal Society University Research
Fellowship and by PPARC.  DW was supported by an EPSRC
Research Studentship.

\end{multicols} 
\end{document}